# Femtosecond Time-resolved MeV Electron Diffraction

Pengfei Zhu[1,2], H. Berger[4], J. Cao[3], J. Geck[4], Y. Hidaka[1], R. Kraus[4], S. Pjerov[1], Y. Shen[1], R.I Tobey[1],
Y. Zhu[1], J.P. Hill[1] & X.J. Wang[1,2*]

[1] *Brookhaven National Laboratory, Upton, NY 11973, USA*
[2] *Key Laboratory for Laser Plasmas & Department of Physics, Shanghai Jiao Tong University, Shanghai 200240, China*
[3] *Physics Dept./NHMFL, Florida State University, Tallahassee, FL 32310, USA*
[4] *IFW, Dresden, Germany*

We report the experimental demonstration of femtosecond electron diffraction using high-brightness MeV electron beams. High-quality, single-shot electron diffraction patterns for both polycrystalline aluminum and single-crystal 1T-$TaS_2$ are obtained utilizing a 5 femto-Coulomb (~$3\times10^4$ electrons) pulse of electrons at 2.8 MeV. The high quality of the electron diffraction patterns confirm that electron beam has a normalized emittance of ~50 nm-rad. The corresponding transverse and longitudinal coherence length are ~11 nm and ~2.5 nm, respectively. The timing jitter between the pump laser and probe electron beam was found to be ~ 100 fs (rms). The temporal resolution is demonstrated by observing the evolution of Bragg and superlattice peaks of 1T-$TaS_2$ following an 800 nm optical pump and was found to be 130 fs. Our results demonstrate the advantages of MeV electron diffraction: such as longer coherent lengths, large scattering cross-section and larger signal-to-noise ratio, and the feasibility of ultimately realizing 10 fs time-resolved electron diffraction.

X-ray Free Electron Laser (FEL) sources and Ultrafast Electron Diffraction (UED) are two of the most powerful tools for exploring the ultrasmall and ultrafast world [1-2]. The large electron scattering cross section and compactness of electron facilities make UED a particularly attractive option for exploring ultrafast processes and the technique has been used for studying strongly correlated electron systems and for mapping the dynamics of bond breaking in gas phase chemical reactions [1-4]. However, to date, the time resolution of such experiments has been limited by pulse broadening from repulsive space-charge effects (SCE) and the limited acceleration field [5].

To reduce such effects, MeV electron beams generated by a photocathode RF gun have been proposed for UED applications [6-8]. In such schemes, a laser is used to illuminate a photocathode, producing a high-brightness electron beam, and to control the initial spatial and temporal distributions of that beam. An RF cavity then rapidly accelerates the electrons to a few MeV. The RF field also compresses the electron beam as it is accelerated in the time-dependent electrical field [6]. For electron pulses at MeV energies, the magnetic field induced by the moving electron beam, together with relativistic effects, greatly reduce the SCE. Specifically, the transverse and longitudinal SCE scale as $1/\beta^2\gamma^3$ and $1/\beta^2\gamma^5$, respectively [9], where $\beta$ and $\gamma$ are the relativistic velocity and energy, respectively. Thus, increasing the electron energy has the potential to compress more electrons into a shorter electron bunch. Another important advantage of relativistic beams is that they eliminate the velocity mismatch between the pump laser and the probe electron beam. This mismatch limits the time resolution of ultrafast dynamics for dilute samples, such as gas and liquid samples [10]. In addition to the higher temporal resolution, MeV electrons can penetrate thicker samples. Finally, the higher electron beam energy leads to a larger elastic scattering cross section and a decrease in the inelastic scattering cross sections [11], increasing the diffraction signal and reducing the sample damage.

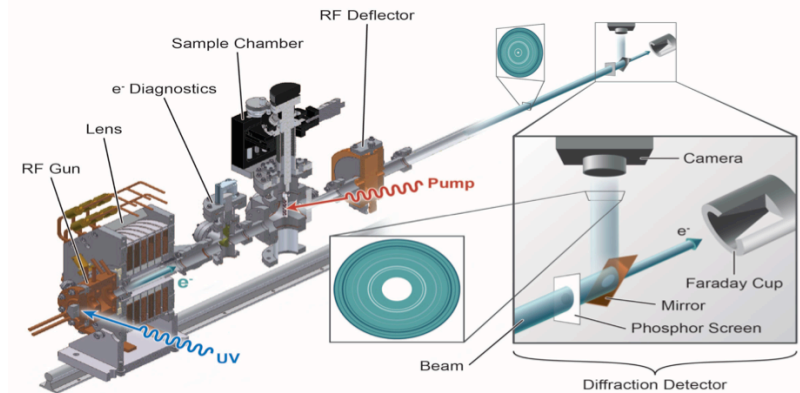

Figure 1: Schematic of the experimental set-up. UV photons from a Ti-sapphire laser are used to generate electrons in the RF gun. A solenoid magnet focuses the beam onto the detector screen 4 m downstream of the sample. Near-IR pulses from the same Ti-sapphire laser are used to optically pump the sample.

However, despite significant efforts in MeV-UED in the last few years [12-15], timing jitter between the pump laser and the MeV electron beam, and the quality of the electron beam, have prevented the ultimate potential of this approach being realized. Here we overcome those challenges and report high-quality, single-shot electron diffraction patterns for both polycrystalline aluminum and single-crystal 1T-TaS$_2$, each obtained utilizing a 5 fC electron bunch (~$3\times10^4$ electrons) at 2.8 MeV. The electron bunch length is estimated to be 40 fs. The timing jitter between the pump laser and the probe electron beam was ~ 100 fs and the overall time resolution, as determined from the evolution of the superlattice peaks of 1T-TaS$_2$ following an 800 nm optical pump, was ~ 130 fs (rms). The signal to noise ratio (SNR) and sensitivity reported here is more than an order of magnitude higher than earlier experiments [12-15]. As a result of these developments, MeV-UED now stands ready to contribute as a powerful probe of ultrafast dynamics, complementing FEL sources with similar sensitivity and temporal characteristics.

To achieve these results, the MeV UED apparatus reported here was carefully optimized at each step, from electron generation to the detection. Furthermore, there is extensive built-in beam instrumentation which is capable of measuring all the basic transverse and longitudinal electron beam parameters, including charge, beam profile and timing jitter. We have determined previously, through simulations, that the optimum electron beam energy is between 2 and 4 MeV [13]. Use of higher electron beam energies requires larger RF fields, which not only introduce nonlinear RF effects that degrade the electron beam quality, but also produce significantly more dark current. The operating energy here is optimized at 2.8 MeV.

A schematic layout of the MeV UED system is shown in Figure 1. The total footprint of the MeV UED system is about 1.5 by 5 meters. The electron source is a BNL (Brookhaven National Lab)-type, 1.6-cell photocathode RF gun [16], capable of producing electron beams with energy up to 5 MeV. The photocathode is the copper back wall of the RF gun cavity. A UV pulse (4.65 eV, ~150 fs (FWHM)) from a frequency-tripled Ti:sapphire laser obliquely incident (~67°) onto the photocathode is used to generate femtosecond electron beams. The radius of the laser spot on the cathode is ~100 μm. The photocathode RF gun is mounted directly on the solenoid magnet to ensure the best alignment between the RF gun and the magnet. The solenoid magnet was optimized for MeV UED, and is followed by a 6-way cross that contains electron beam diagnostics, consisting of a movable Faraday cup and a beam profile monitor. A motorized, 1 mm diameter collimator is also installed on the 6-way cross. The sample chamber is equipped with an x-y-Φ motorized manipulator with cryogenic cooling capability. A window on the sample chamber, located 18° off the beam axis, is used to bring the pump laser in, as well as for monitoring the electron beam and pump laser positions. The sample holder can carry up to 9 different specimens, plus one YAG screen used for spatial alignment of the pump laser, electron beam and the sample. An RF deflecting cavity is installed immediately after the sample chamber. The electron diffraction detector is positioned four meters downstream of the sample chamber to optimize reciprocal space resolution.

One of the challenges in realizing MeV UED has been the diffraction detector. Traditional UED detection techniques, such as a fiber coupler, are not applicable at MeV energies. Furthermore, the x-ray background produced by the MeV electron beam must be carefully avoided. Here, we utilize a 4-cm-diameter phosphor screen, positioned perpendicular to the electron beam. The phosphor type, grain size and thickness were all selected to maximize the number of photons generated by the MeV electron beam. A 45° copper mirror with a 5 mm hole is placed behind the phosphor screen. The hole allows the non-diffracted electrons to pass through; a Faraday cup positioned immediately following the copper mirror acts as a beam stop. An Andor electron-multiplying CCD (EMCCD) camera with an f#1.4 large-aperture lens records the image reflected by the mirror. The phosphor screen and copper mirror are mounted on an actuator so that they can be retracted and the charge of the all electrons interacting with the sample can be measured accurately.

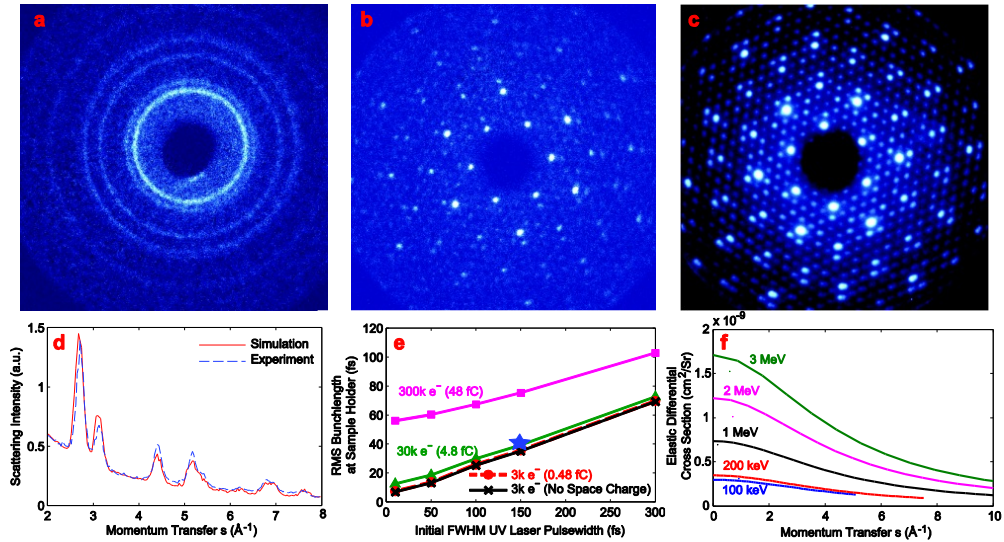

**Figure 2** Single-shot diffraction pattern from (a) polycrystalline aluminum and (b) 1T-TaS$_2$. (c) 100-shot diffraction pattern from 1T-TaS$_2$ (d) Simulated and experimental scattering intensity as a function of momentum transfer, s, for polycrystalline Al. (e) Simulated bunch lengths (rms) at the sample position as a function of initial UV laser pulse widths (FWHM) for various electron bunch charges. The blue star represents the experimental working point, corresponding to a 40 fs bunch length. (f) The elastic electron scattering differential cross sections from Al as a function of momentum transfer for various electron beam energies.

The UV and pump laser optics are installed on optical tables next to the UED system with a one-foot-long motorized delay stage on the pump optics arm.

The Ti:sapphire laser system consists of an RF synchronized oscillator, a regenerative amplifier and a two-pass amplifier. The oscillator is synchronized to an RF master oscillator at 81.6 MHz, and has a measured time jitter of ~50 fs relative to the RF system when the oscillator environment is controlled to within ± 0.1 °F. The equipment is installed at the BNL Source Development Laboratory (SDL) [17].

The electron beam is first optimized by adjusting the relative phase between the UV laser and the RF gun, to about 15 °. This value was found to optimize the quality of the electron beam and is consistent with simulations and previous experimental results [6]. Then by focusing the electrons onto the diffraction detector, convergent-beam diffraction is employed. The electron beam energy was optimized to be ~ 2.8 MeV by observing the diffraction pattern of the aluminum film.

To demonstrate the quality and time resolution of the MeV UED system, two well-understood samples were employed. The first sample was a 100-nm-thick polycrystalline aluminum film; the second, a single-crystal 1T-TaS$_2$ film with a charge density wave (CDW) modulation. Figures 2a and 2b are single-shot, electron diffraction patterns for the two samples each generated by a single 5 fC ($3\times10^4$ electrons) electron pulse. The previous record for single-shot femtosecond diffraction was with 200 fC pulses on a gold film [18]. Here we have used a factor of 40 fewer electrons in a single pulse. In fact, the increase in sensitivity is even larger than this, when the difference in atomic number (Al vs. Au) in the two experiments is taken into consideration. This is another factor of approximately $(Z_{Au}/Z_{Al})^{1.5} = (79/13)^{1.5} \approx 15$. Our experimental results thus show MeV electron diffraction has a higher sensitivity than conventional energy diffraction, in contrast to the prevalent view [18].

It is worth emphasizing this latter point. The improved sensitivity comes from two sources. First, in the small scattering angle regime (< 10 mrad) the elastic scattering differential cross sections increases with electron energy, as shown in figure 2f. In the present set up, the acceptance angle is about 5 mrad corresponding to a maximum momentum transfer of 8.3 Å$^{-1}$ for the 2.8 MeV electron beam. Secondly, the higher electron energy allows the use of a thicker phosphor screen. This will generate visible photons more efficiently: An MeV electron deposits about 17 keV energy when passing though a 30 $\mu$m thick phosphor screen. Since the phosphor conversion efficiency is ~20 eV per photon, this yields about 850 photons per electron [19].

In Figure 2b, single-shot diffraction from 1T-TaS$_2$ is shown. Both the Bragg peaks and the CDW superlattice peaks are clearly visible with good SNR. We believe that this is the first single-shot image showing superlattice peaks with a femtosecond electron beam. Figure 2c is a 100-shot accumulation of the 1T-TaS$_2$ diffraction pattern, demonstrating a significant improvement in quality compared to the previous ultrafast electron diffraction using a DC gun [18]. The visibility of the superlattice peaks demonstrates not only the good transverse electron beam quality, but also the good overall SNR (> 400) of the system.

To quantify the electron beam quality, we performed start-to-end simulations of diffraction from an Al film using the computer code - General Particle Tracer (GPT) [20]. Figure 2d shows the simulated scattering intensity as a function of momentum transfer, *s*, together with the experimental pattern, as obtained by accumulating 100 shots and azimuthally averaging the resulting pattern. The good agreement between the simulation and the experiment confirms the validity of the simulation. This allows us to extract a normalized transverse beam emittance of ~50 nm-rad (rms), which is dominated by the thermal emittance. The transverse coherent length on the sample is estimated to be ~11 nm, which is comparable to that achieved with the low energy femtosecond *single-electron* pulses [ 21]. Using the same operating parameters, we also extracted the electron beam energy spread and find it to be ~ 0.02%, which corresponds to a longitudinal coherence length of ~2.5 nm.

In figure 2e, the electron bunch length is shown as a function of the UV laser pulse width for various different electron bunch charges, as calculated with the same parameters used in the simulation in Figure 2d. One sees that for a bunch charge of 5 fC or less, there is negligible bunch lengthening from the space charge effects. This implies that the bunch length is about 40 fs (rms) for the 150 fs (FWHM) laser pulse length used here. These simulations also indicate that the MeV UED electron bunch length could be as short as 10 fs if a 20 fs commercial laser system is used.

We measure the electron beam temporal distribution of the present set up by utilizing the RF deflector as a streak camera [22], i.e. converting temporal information into spatial information (Fig. 3a). Since the RF deflector is powered from the same klystron as the RF gun, there is an inherent synchronization between the two. The RF deflector can be calibrated by monitoring the centroid displacement as the RF deflector phase is adjusted (Fig. 3b). For the S-band (2.856 GHz) RF system employed here, the calibration is 1 RF degree = 0.97 ps ≈ 9 pixels (Fig. 3b). The electron beam bunchlength was characterized by observing its spot size change without and with the deflector on. Due to the resolution limit, we can only conclude that the electron beam bunch length is shorter than 100 fs, which is consistent with our estimation of 40 fs, as seen in the simulations.

The electron beam arrival time jitter can be determined by measuring the spread of the electron beam centroid on the diffraction detector. As a result of positional fluctuations due to the laser pointing stability and RF fluctuations, this spread will have a finite width even if the RF deflector is turned off. When the deflector is turned on, the width of this spread increases because the timing jitter is now converted into spatial jitter and added to the original positional jitter. The difference in these two distribution widths provides a measurement of the timing jitter. From the data of Figure 3c, we estimate the timing jitter is ~ 100 fs (rms) as seen by the broadening of the curves.

To demonstrate the overall temporal resolution of our MeV UED, we performed pump-probe experiments on the CDW material 1T-TaS$_2$. Figure 4a shows the temporal evolution of the intensity of the Bragg and superlattice peaks following the 800-nm, 150-fs pump laser pulses. The pump laser power density is ~1.5 mJ/cm$^2$. We observed similar behavior to that reported in an earlier study[23]. Specifically, the atomic motions that are driven by the optically induced change in the electronic energy distribution result in the suppression of the charge density wave state. Figure 4b shows the temporal behavior of the superlattice peak, together with a fit to a Gaussian experimental resolution convolved with the reported time constant of 300 fs for this material [23], where we have modeled the sample behavior as a simple exponential decay function. We find the best fit for an experimental time resolution of 130 fs (rms). Since the electron bunch length is short (40 fs), this time resolution is dominated by two sources: the timing jitter between the electron beam and pump laser, and pulse length of the pump laser. With state-of-the-art RF and laser technologies, such as solid state klystron modulators and a diode pumped laser system, we anticipate that this time resolution could be improved by an order of magnitude, to about 10 fs.

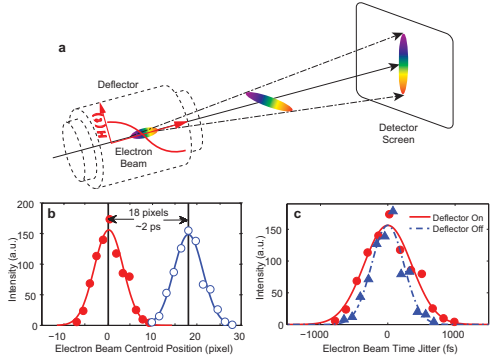

Figure 3: (a) Schematic of RF deflecting cavity for femtosecond bunch timing measurements. (b) Calibration of the RF cavity; a shift in phase of 2° causes a positional shift of 18 pixels. (c) With the RF cavity on, timing jitter is converted to spatial jitter and thus contributes additional width to the distribution. The timing jitter is thus estimated to be 100 fs.

In summary, we report high quality, single-shot femto-second electron diffraction with a 5 fC electron bunch at 2.8 MeV. Our experimental results show that MeV UED not only can achieve the highest temporal resolution (here 130 fs), but also has significantly improved sensitivity compared to the existing methods. The optimized MeV electron beam in our system, including large transverse and longitudinal coherence lengths, allows us to record weak superlattice reflections with unprecedented quality. Such charge-orbital-order related reflections sometimes can be difficult to detect, especially when the crystals are small [24]. Continued improvement of such a UED system, such as solid-state klystron modulator and a state-of-art laser system, could improve the MeV UED temporal resolution to 10 fs. By reducing the laser spot size (from 100 $\mu$m to 10 $\mu$m) and the thermal emittance, the transverse coherence length could be as large as 100 nm, this may lead to electron diffraction-based coherent diffractive imaging, or ptychography, to visualize aperiodic nano-objects in real space, including nanocrystals, proteins and molecules, with a high temporal time resolution. Thus MeV UED stands ready as a technique to complement free electron laser x-ray sources in the study of ultrafast structure dynamics for a wide range of systems.

The authors would like to thank H. Ihee, C.C. Kao, J. Misewich and J.B. Murphy for discussions and encouragement. The technical support by National Synchrotron Light Source (NSLS) and BNL Photon Science Directorates is gratefully acknowledged. This research is supported in part by the US Department of Energy under Contract No:DE-AC02-98CH1-886, and BNL Laboratory Directed Research and Development (LDRD) funds 2010-010, LDRD 2012-22, and China Natural Science Foundation grant No. 113279031.

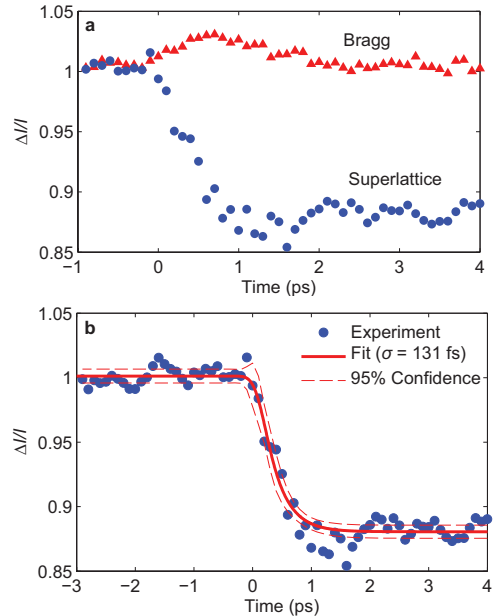

Figure 4: Characterization of time resolution for MeV UED. (a) Dynamics of Bragg and superlattice peaks of 1T-TaS2. (b)The overall time resolution of the instrument is estimated to 130 fs by fitting the superlattice data to an exponential decay.

* Corresponding author: xwang@bnl.gov

[1] M. Chergui and A. H. Zewail, ChemPhysChem **10**, 28 (2009).
[2] G. Sciaini and R. J. D. Miller, Rep. Prog. Phys. **74**, 096101 (2011).
[3] G. Mourou and S. Williamson, Appl. Phys. Lett. **41**, 44 (1982).
[4] C. Hensley *et al*, Phys. Rev. Lett. **109**, 133202(2012).

[5] X.Wang, *et al*, Rev. Sci. Instrum. **80**, 013902(2009).
[6] X.J. Wang, et al., Phys. Rev. E 54, R3121 (1996).
[7] X.J. Wang *et al*, Proce of 2003 Part. Accel. Conf. 420(2003).
[8] X.J. Wang *et al*, J. Korean Phys. Soc. **48,** 390 (2006).
[9] M. Reiser, Theory and Design of Charged Particle Beams, John Wiley and Sons, Inc., New York, 1994.
[10] P. Reckenthaeler *et al*. Phys. Rev. Lett. **102**, 213001 (2009).
[11] D.K. Bhattacharya and N.N.DAS GuPTA, Ultramicroscopy 5, 75-79 (1980).
[12] J.B. Hasting *et al*. Appl. Phys. Lett. **89**, 184109 (2006).
[13] R.K. Li *et al*.,Rev. Sci. Instrum. **80**, 083303 (2009).
[14] P. Musumeci *et al*., Appl. Phys. Lett. **97**, 063502(2010).
[15] Y. Murooka *et al*, Appl. Phys. Lett. **98**, 251903 (2011).
[16] K. Batchelor *et al*, Nucl. Instrum. Methods Phys. Res. A 318, 372 (1992).
[17] J.B. Murphyand X.J. Wang, Synchrotron Radiat. News **21**, 41(2008).
[18] T. van Oudheusden *et al.* Phys. Rev. Lett. **105**, 264801 (2010).
[19] D.P. Russell, Thesis (PH.D.), PRINCETON UNIVERSITY(1992).
[20] http://www.pulsar.nl/gpt/.
[21] F. Kirchner *et al*, New Journal of Physics 15, 063021 (2013).
[22] X.J. Wang *et al*, Nucl. Instrum. Methods Phys. Res. A 356, 159 (1995).
[23] M. Eichberger et al, Nature 468, 799–802 (2010).
[24] L. Piazza, A. Mann, C. Ma, H.Yang, Y. Zhu, L. Li, and F. Carbone, Structure Dynamics, 1 014501 (2014).